\documentclass[useAMS,usenatbib,dcolumn,usegraphicx,twocolumn,10pt]{mnarxiv}
\usepackage{amssymb}
\usepackage{float}
\usepackage{placeins}
\usepackage{xspace}
\usepackage{natbib}
\usepackage{hyperref}

\newcommand{\ud}[1]{\mathrm{d}#1}
\newcommand{\br}[1]{\left(#1\right)}
\newcommand{\pc}{\mathrm{pc}}
\newcommand{\kpc}{\mathrm{kpc}}
\newcommand{\kms}{\mathrm{km}/\mathrm{s}}
\newcommand{\A}{\textsf{A}\xspace}
\newcommand{\B}{\textsf{B}\xspace}
\renewcommand{\exp}[1]{\mathrm{e}^{#1}}
\renewcommand{\sun}{_{\circ}}
\newcommand{\obj}{_{\otimes}}

\newcommand{\abs}[1]{\left|#1\right|}
\renewcommand{\sq}[1]{\left[#1\right]}
\newcommand{\dd}[1]{\mathrm{d}{#1}}
\newcommand{\msun}{M_{\odot}}
\newcommand{\lsun}{L_{\odot}}
\newcommand{\ndm}{\textsf{NDM}\xspace}

\newcommand{\eqref}[1]{(\ref{#1})}

\title[Microlensing...]{Gravitational microlensing as a test of a finite-width disk model of the Galaxy}
\author[]{{Szymon Sikora$^{1}$},
{{\L}ukasz Bratek$^{2}$},
{Joanna Ja{\l}ocha$^{2}$},
{Marek Kutschera$^{1}$}
\\
$^{1}$Institute of
Physics, Jagellonian University,  Reymonta 4, PL-30059 Krak{\'o}w, Poland\\
$^{2}$Institute of Nuclear Physics,
Polish Academy of Sciences, Radzikowskego 152, PL-31342 Krak\'{o}w, Poland}

\begin{document}
\date{\today}
\pagerange{\pageref{firstpage}--\pageref{lastpage}} \pubyear{2009}

\maketitle

\begin{abstract}
The aim of this work is to show, in the framework of a simple finite-width disk model, that the amount of mass seen through gravitational microlensing measurements in the region $0<R<R\sun$ is consistent with the dynamical mass ascertained from Galaxy rotation after subtracting gas contribution. Since microlensing only detects compact objects, this result suggests that a non-baryonic mass component may be negligible in this region.
\medskip
\hrule
\flushleft \textbf{The definitive version is available at\\
\url{http://dx.doi.org/10.1051/0004-6361/201219926}}
\medskip
\hrule
\end{abstract}


\section{Introduction}

Spiral galaxies are customarily regarded as disk-like objects imbedded
in a spheroidal halo of non-baryonic dark matter (\ndm).
In this picture the rotation of galaxies is governed mainly by the gravitational field of \ndm halo, at least at larger radii.

But this scenario might not be typical, because rotation of some of spiral galaxies can be explained with a negligible (if any) amount of \ndm.
For these galaxies, a  thin disk model approximation has proven to be quite satisfactory
\citep{2008ApJ...679..373J,2010MNRAS.406.2805J}.
There was also a successful attempt at explaining high transverse gradients in the Galaxy rotation in the same disk model framework \citep{2010MNRAS.407.1689J}, whereas including a significant spheroidal mass component proved to reduce the predicted gradient to values that are inconsistent with measurements.
There is also a recent result \citep{Bidin} showing that the kinematics of the thick disk stars in the solar neighborhood is consistent with the visible mass alone, and there is no place for a spheroidal
 distribution of NDM, although the assumptions of this result are the subject of debate \citep{2012arXiv1205.4033B}.

The above results suggest that the overall mass distribution in Galaxy might be flattened rather than spheroidal, at least in the Galaxy interior.
In this case,  the disk model description would be more appropriate to describing the true mass distribution in this region than mass models with a dominating spheroidal mass component.
Further independent arguments that would help for deciding about the properties of mass distribution in Galaxy are necessary.
To this end we apply the gravitational microlensing method as a means to determining the mass residing in compact objects, and compare the results for radii $R<R\sun$ with predictions of a finite-width disk model.

\begin{figure}\centering
\includegraphics[width=0.5\textwidth]{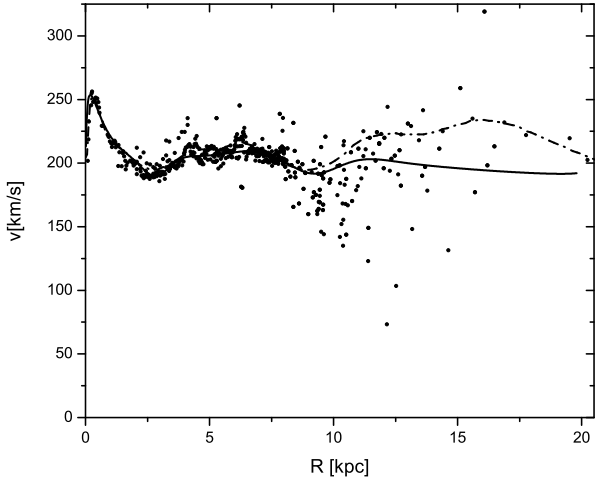}
\caption{\label{fig:drogarotacje}Rotation of Galaxy. Unified rotation velocity measurements [\textit{points}] (see the text for references)  and two model rotation curves \A  [\textit{thick line}] \citep{2009PASJ...61..227S} and \B  [\textit{dot-dashed line}] \citep{1999ApJ...523..136S} used to obtain surface mass density.   }
\end{figure}

It is important that in the same region where microlensing results are known, the Galactic rotation curve is relatively well determined. Figure
\ref{fig:drogarotacje}
shows rotation data for the Galaxy from various independent measurements
\citep{1978A&A....63....7B,1985ApJ...295..422C,1989ApJ...342..272F,
2007A&A...473..143D,1982ApJS...49..183B,1997PASJ...49..453H},
together with two model rotation curve fits obtained in \citep{2009PASJ...61..227S} and \citep{1999ApJ...523..136S}, which we use in our calculations and refer to as \A and \B, respectively. They are used, because they agree well with the data points in the region of interest. Rotation curve \A was obtained assuming the reference velocity $\Theta_{\circ}=200\kms$. Rotation curve \B
assumes $\Theta_{\circ}=220\kms$.
Our results are based on these two curves, although in sec.\ref{section_results}, in order to estimate the influence of the uncertainty in the determination of the parameter $\Theta_{\circ}$ on our results concerning the optical depth,
we also use a rotation curve from \citep{2011MNRAS.414.2446M} that is based on the terminal rotation curve data and a high value $\Theta_{\circ}=239\kms$ preferred by the author.

\begin{figure}\centering
\includegraphics[width=0.5\textwidth]{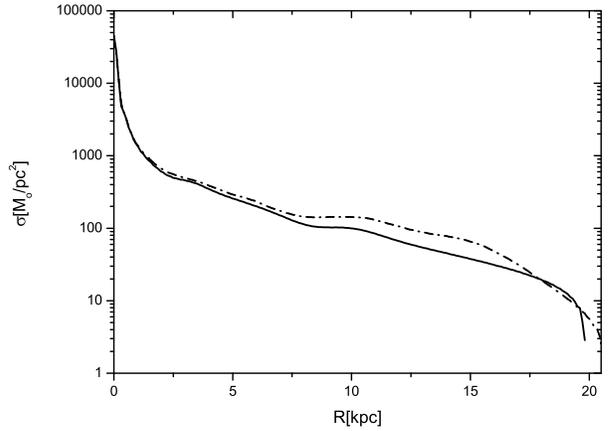}
\caption{\label{fig:drogagestosci}Surface mass densities in the thin disk model corresponding to rotation curves \A [\textit{solid line}] and \B  [\textit{dot-dashed line}]. }
\end{figure}

The surface mass density in thin disk model can be obtained with the help of the following integral mapping $v(R)$ -- the tangential component  of Galaxy rotation in the disk plane -- to the surface mass density $\sigma(R)$ in the disk plane (representing a column mass density):
\begin{equation}\label{eq:sigmamoja}\sigma\br{R}=\frac{1}{2\pi^2G}\int\limits_{0}^{\infty}
\sq{
\frac{K\!\br{\mu_x}}{1+x}-
\frac{E\!\br{\mu_x}}{1-x}}\frac{v^2\!\br{R\,x}}{R\,x}\,\ud{x}.\end{equation}
Here,
 $R$ is the radial variable in the disk plane, $K$ and $E$ are complete elliptic integrals of the first and second kind,\footnote{Since there are two conventions for the argument of elliptic integrals met in the literature,  we give the definitions which we use \citep{Ryzhik}:\[
K(k)=\int\limits_{0}^{\pi/2}\frac{\ud{\phi}}{
\sqrt{1-k^2\sin^2{\phi}}},\qquad
E(k)=\int\limits_{0}^{\pi/2}\ud{\phi}
\sqrt{1-k^2\sin^2{\phi}}\] } $\mu_x=\frac{2\sqrt{x}}{1+x}$ and $x$ is a dimensionless integration variable.
Mathematical details are given in Appendix \ref{appendix_A}. In studying galaxies with the use of the expression \eqref{eq:sigmamoja}, it is assumed that a galaxy is a flattened disk-like object rather than spheroidal  and that the motion of matter is predominantly circular. The central bulge  can also be described in this model by the appropriate substitute column mass density.  One should recall that the relation between the mass distribution and the rotation law is qualitatively  different from that valid under spherical symmetry. It is nonlocal in the sense that the total rotation field is needed to determine the surface mass density at a given radius. Cutting off integration introduces some error, however it is usually not significant for radii smaller than $0.6$ of the cutoff radius \citep{2008MNRAS.391.1373B}. This condition is satisfied in the region of our interest. The resulting surface mass densities corresponding to rotation curves \A and \B are shown in figure  \ref{fig:drogagestosci}.

\section{The microlensing method}
In contrast to methods based on analyzing the Galactic rotation or the motions of halo objects, the microlensing method gives us the opportunity of measuring Galaxy mass independently of Galaxy dynamics. The method was developed and described in detail by several authors \citep{1986ApJ...304....1P}, \citep{1996ARA&A..34..419P}, \citep{schneider2006gravitational}. To estimate the integrated mass distribution along various lines of sight, the method employs the phenomenon of gravitational light deflection of distant sources of light in the field of point-like masses distributed in space.
 A directly measured quantity in this method is the optical depth.  Before going to the main analysis, let us discuss this important notion in more detail.

\subsection{Optical depth along a line of sight}
Suppose a point-like lens of mass $M$ is located exactly on a line of sight between the source of light and the observer. This line  plays the role of the optical axis. Then, the image of the source forms the Einstein ring of some angular radius  $\theta_E$ as perceived by the observer and concentric with the optical axis.
Let $D_L$ and $D_S$ be the distances along the optical axis between the observer and the lens and between  the observer and the source, respectively.
Expressed in terms of $\theta_E$ and the gravitational deflection angle $\alpha_E$, the lens equation in the weak field approximation reduces in this particular configuration to $\theta_E\,D_S=\alpha_E (D_S-D_L)$. Taking into account that  $\alpha_E=\frac{4GM}{c^2 R_E}$ with $R_E=\theta_E D_L$, we obtain
$
\theta_E^2=\frac{4GM}{c^2D_L}\br{1-\frac{D_L}{D_S}}
$.

The {optical depth} is by definition the probability of finding a compact object (a lens) within its Einstein radius on some plane perpendicular to the line of sight. In a given volume element $\ud{S}\,\ud{D_L}$, large enough to contain many lenses of various masses, the corresponding probability is a sum $\sum_i\nu_i\pi R_{E_i}^2\ud{D_L}$ over all lenses of mass $M_i$, where $\nu_i$ is the number density of lenses of mass $M_i$. The ratio $R_{E_i}^2/M_i$ is independent of $M_i$, and $\sum_i\nu_iM_i$ is simply the volume mass density of compact objects (lenses) $\rho_{co}$. Integration over $D_L$ along a given optical axis therefore gives us  $$
\tau=\int\limits_{0}^{D_S}\frac{4\pi G}{c^2D_L}\br{1-\frac{D_L}{D_S}}\rho_{co}\br{\vec{r}(D_L)}D_L^2\ud{D_L}.
$$ This is the (integrated) optical depth along the line of sight.

For practical reasons, the passage of a lens in front of a source of  light is assumed to have occurred when  the magnification of the source exceeds some threshold value. The optical depth is measured by means of counting the number of such events for millions of stars located near the Galactic center. A comprehensive description of such a measurement can be found in \citet{2010GReGr..42.2047M}.

\subsection{\label{sec:appl}Application in the framework of a finite-width disk model}

A thin disk model predicts some column density $\sigma$ from Galaxy rotation data with the help of the formula discussed earlier. To compare the disk model predictions with the optical depth measurements, an appropriate volume mass density entering the integral defining $\tau$ is needed. In the disk approximation, $\sigma$ can be regarded as a column mass density assigned to a volume mass density $\rho$. This identification is also frequently made in modeling galactic disks. The column mass density is
defined by means of the integral
$
\sigma(R)=\int_{-\infty}^{\infty}\,\rho(R,Z)\,dZ
$, where $R,Z$ are Galacto-centric cylindrical coordinates and the axial symmetry has been assumed.

The integration in  $\tau$ is carried out along a line of sight joining the observer located at $X_\odot=\sq{R\sun,0,0}$ and a source located at $X_{\otimes}$, where
$X_{\otimes}=X_\odot+(1+\chi)R\sun \sq{-\cos{b}\cos{l},-\cos{b}\sin{l},\sin{b}}$.
Here, $b$ is the Galactic latitude, $l$ the Galactic longitude, and  $(1+\chi)R\sun$ is the distance of the source of light from the Sun ($\chi$ is a dimensionless position parameter). A convenient parametric description of the line of sight is $X(s)=X_\odot+s (X_{\otimes}-X_\odot)$ (then $D_L(s)=s(1+\chi)R\sun$ and $D_{S}=(1+\chi)R\sun$). On assuming the standard exponential profile $
\rho(R,Z)=\rho(R,0)\,\exp{-\abs{Z}/h}
$ with a fixed scale height $h$,  the line integral  $\tau$ becomes a function  of $b$, $l$ and $\chi$:
\begin{eqnarray}
\label{eq:depth_full}
&\tau_h(l,b,\chi)=\frac{2\pi G R\sun^2}{c^2h}\int\limits_0^1 \br{1+\chi}^2\exp{-s(1+\chi)\abs{\sin{b}}R\sun/h}\times\\
&\sigma\br{R\sun \sqrt{1+(1+\chi)s\cos{b}\sq{(1+\chi)s\cos{b}-2\cos{l}\,}}} s(1-s)\ud{s}\nonumber
\end{eqnarray}
where
$\sigma$  is a known function.

The observed sources of light are located in the vicinity of the Galactic center.  The measurements we use are presented as pairs $(\tau,b)$ for various $b$ with some errors $\Delta\tau$. They are taken from \citep{2010GReGr..42.2047M} as averages over a domain of $l$ around Galactic center. All of the fields used are placed within the area presented in the figure 14 in \cite{2006A&A...454..185H}, although the most important region is $l\in(-5^\circ,5^\circ)$. The precise value of $\chi$ is not known and its uncertainty must somehow be taken into account in the model of $\tau$. Either way, in a first approximation one can assume $l=0$ and $\chi=0$, in which case \eqref{eq:depth_full} reduces to $\tau_h(0,b,0)\approx\tau_h(b)$, where
 \begin{equation}\label{eq:depth}
\tau_{h}(b)=\frac{2\pi G R\sun^2}{c^2\,h}\int\limits_{0}^{1}{\sigma\br{R\sun\br{1-s}}}
\,\exp{-s\abs{b}R\sun/{h}}s\br{1-s}\ud{s}.
\end{equation}
In $\tau_{h}(b)$ we used the approximations $\cos{b}\approx 1$ and $\sin{b}\approx b$, since $\abs{b}<0.1$ radians for the observed sources of light.

Before using the above integrals, a remark should be made concerning the appropriate choice for $\sigma$. We recall that the disk model  gives the surface density describing the gravitating masses, regardless of their nature, whereas  the microlensing events only concern compact objects. Accordingly, in making comparisons, the contribution from continuous media such as gas should be subtracted from the density.

\begin{figure}\centering
\includegraphics[width=0.5\textwidth]{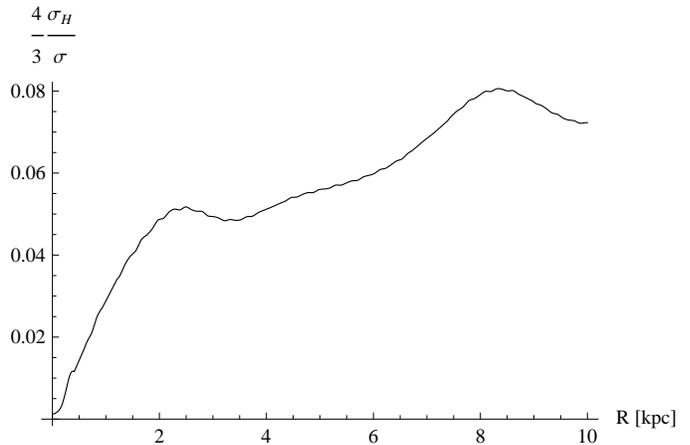}
\caption{The ratio of the hydrogen surface mass density $\sigma_H$ relative to the total surface density accounting for the Galaxy rotation in the thin disk model. A factor $4/3$ has been included to account for the helium abundance.\label{sigma_h}}
\end{figure}

For estimating the gas contribution  we used the neutral and molecular hydrogen density profiles obtained by \citet{2006A&A...459..113M} from COBE/DIRBE and COBE/FIRAS observations. The column mass density of gas is then found to be approximately four third of that for hydrogen. It includes the contribution from helium. The result of this procedure is shown in figure \ref{sigma_h}.

\begin{figure}\centering
\includegraphics[angle=-90,width=0.5\textwidth]{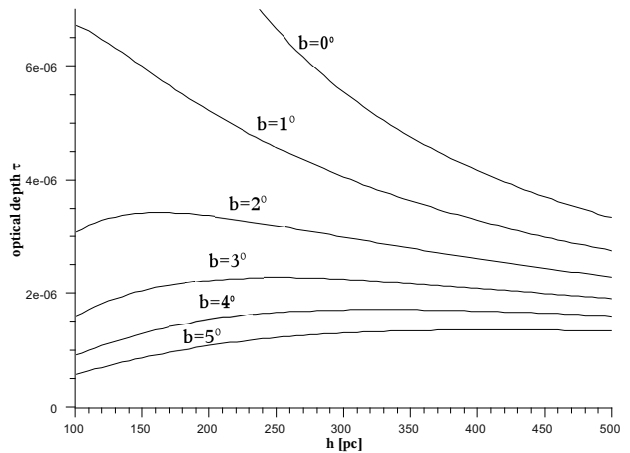}
\caption{\label{tau_h} The optical depth $\tau_h(b)$ in disk model as a function of the disk scale height $h$ shown for various lines of sight, labeled by the corresponding latitudes $b$ (in degrees).}
\end{figure}

\noindent
The relation $\tau_h(b)$ is plotted in figure \ref{tau_h} for several values of  $b$.  As is seen, the predicted optical depth has a correct order of magnitude. Its dependence on the scale height is significantly weaker at higher latitudes.

\noindent
To find a better model of the optical depth, the uncertainty in $\chi$ in the integral \eqref{eq:depth_full} and an averaging procedure of the observational data over $l$ must be simulated. To this end a {Monte Carlo}
method can be used.

\subsubsection{A Monte Carlo method}
 For a given value of $b$, a collection of lines of sight (called ensemble) is chosen randomly with some probability distribution for various pairs $(l,\chi)$, so that $X\obj$ is
 always in some region in the vicinity of the Galactic center. Then $\tau\br{l,b,\chi}$ is calculated for every element of the ensemble, and next, the average value over the ensemble and the respective dispersion is determined. In our simulation we assumed a uniform distribution for $\chi$ and $l$: $\abs{\chi}<0.125$ (that is, $\pm1\kpc$), $|l|<0.017$ (that is, $\pm1^{\circ}$)\footnote{In the simulation that also takes the bar into account, the full range $l=\pm5^\circ$ is considered (see section \ref{section_bar})}. We also used an auxiliary uniform random variable $\Sigma\in(0,\sigma(0))$. At a fixed $b$ we choose those from among the random triples $(\chi,l,\Sigma)$
 for which $\Sigma<\sigma(R)$.
 As a result, the distribution of $X\obj$ is weighted by $\sigma$.

\subsection{The results}\label{section_results}

The frequently used value for the scale height is $h=325\,\pc$. This standard choice is consistent with the  scale height parameter for M-dwarfs of $320\pm50\,\pc$ obtained  by \citet{1997ApJ...482..913G}.

\begin{figure}\centering
\includegraphics[angle=-90,width=0.5\textwidth]{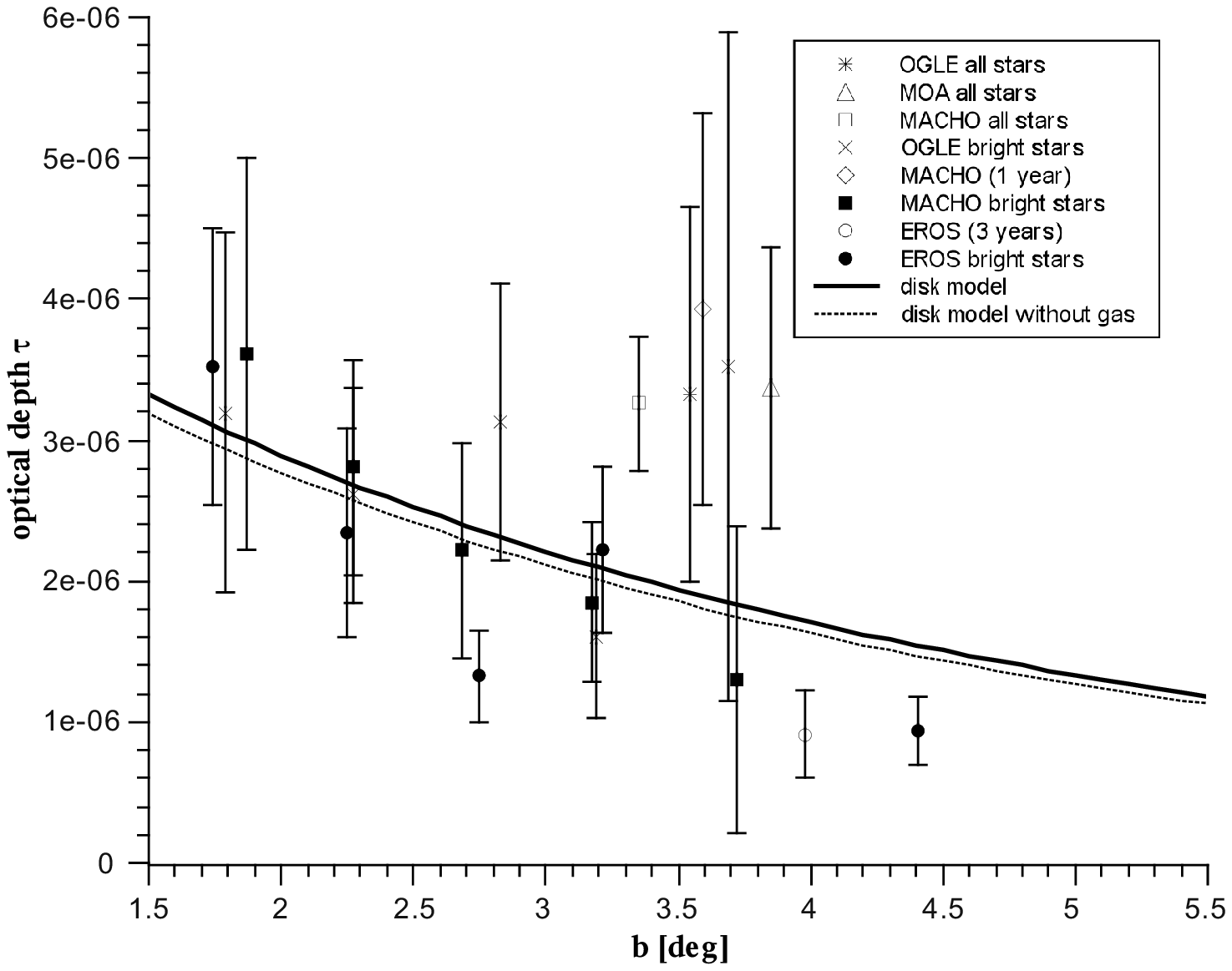}
\includegraphics[angle=-90,width=0.5\textwidth]{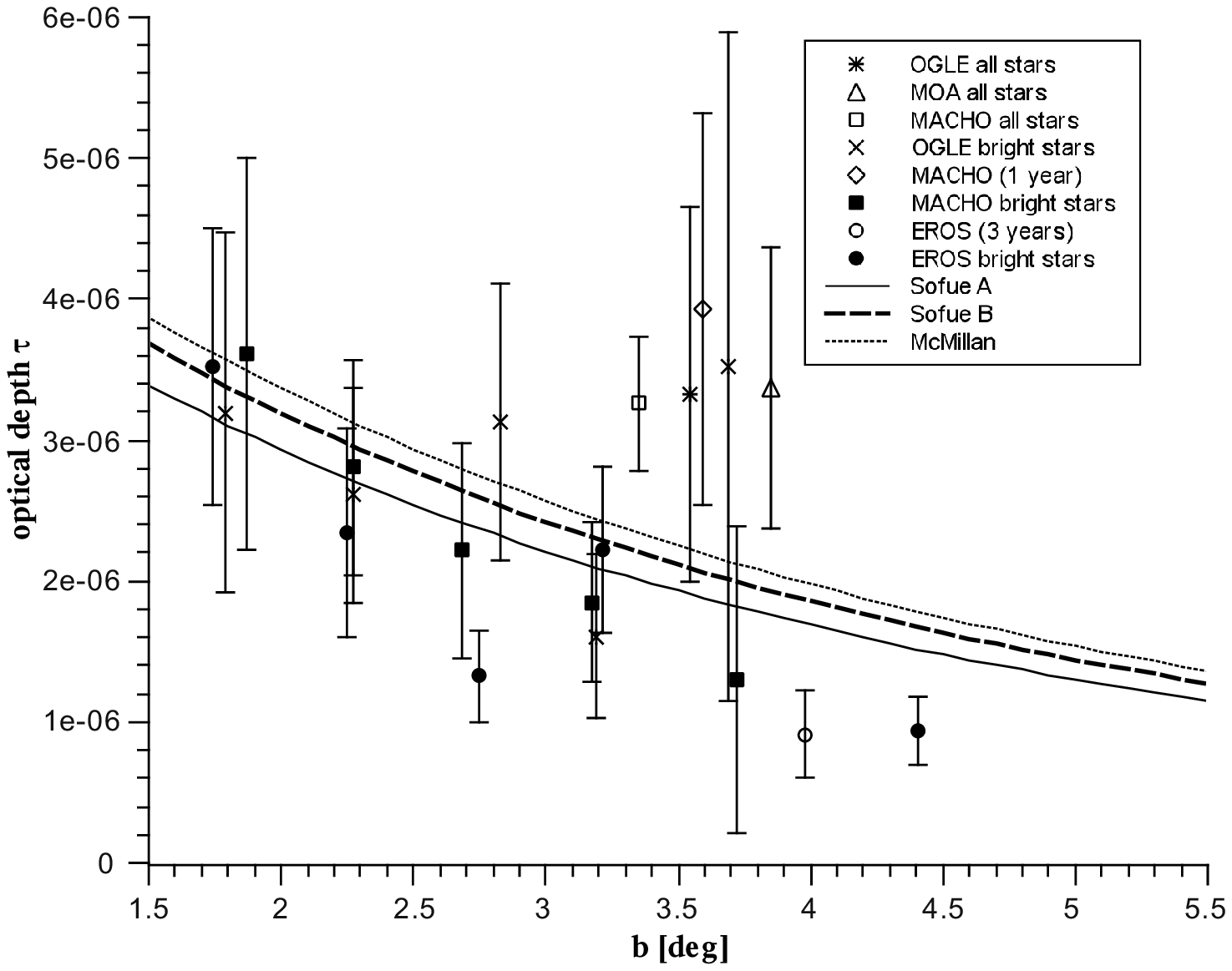}
\caption{\label{tau_b_h325} Optical depths in disk model with the disk scale height $h=325\,\pc$. Points represent the observational data averaged over longitude within the range out to $l=\pm5^\circ$.
 \textsc{Top panel}: The result for the rotation curve \A [\textit{thick line}] and after subtracting the gas contribution [\textit{thin line}]. \textsc{Bottom panel}: Comparison of optical depths for different model rotation curves: \A [\textit{solid line}] and \B [\textit{dashed line}] and, in addition, the rotation curve based on the McMillan data [\textit{dotted line}] (without gas subtraction and with $R\sun=8.29\kpc$ preferred by McMillan).}
\end{figure}

\begin{figure}\centering
\includegraphics[angle=-90,width=0.5\textwidth]{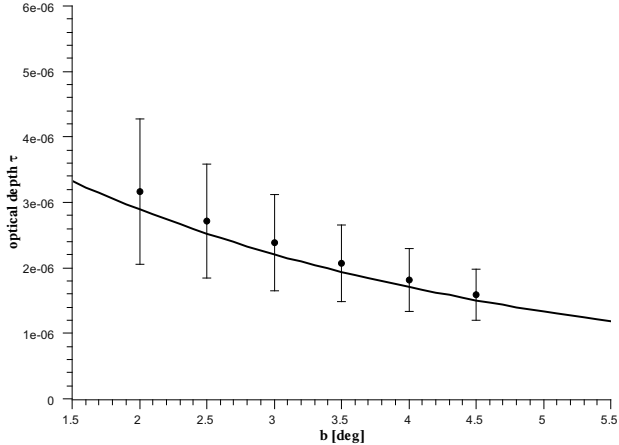}
\caption{\label{fig:MonteCarlo} A Monte Carlo simulation of the optical depth (and standard deviation) at various latitudes $b$ [\textit{solid circles+bars}],  compared with another model (discussed earlier) assuming the sources of light to be located on the symmetry axis [\textit{solid line}].}
\end{figure}

The optical depth calculated in accord with equation \eqref{eq:depth} is plotted in figure \ref{tau_b_h325} for $h=325\,\pc$ and compared with   the {optical depths}
 measured by several collaborations: MACHO \citep{2005ApJ...631..879P}, OGLE \citep{2006ApJ...636..240S}, EROS \citep{2006A&A...454..185H}, and MOA \citep{2003ApJ...591..204S}. These data were collected and discussed in a review \citep{2010GReGr..42.2047M}. There is a subclass in the data consisting of  bright stars. Separation of bright sources allows reducing the blending effect, which is a serious source of uncertainty in the optical depth  \citep{1997A&A...321..424A}, \citep{2007MNRAS.380..805S}. It is seen that the optical depth predicted in the framework of our model fits the data quite well  when the bright stars are considered. In figure \ref{fig:MonteCarlo} the result of the Monte Carlo procedure described earlier was shown and compared with the result shown in figure \ref{tau_b_h325}.
 The simulation reproduces the optical depths consistently with measurements. The Monte Carlo standard deviation also takes some information about the expected uncertainty into account in the distribution of sources around the Galactic center (as a function of longitude and distance to the source $D_S$). It is worth noting that this uncertainty is comparable with the observational error bars.
Moreover, comparing results in  fig.\ref{tau_b_h325}, shows that determining the optical depth hardly depends on the assumed rotation curve, suggesting indirectly that the result is not very sensitive to changes in parameter $\Theta\sun$ in the range from $200\kms$ to $239\kms$. For comparison, in addition to the rotation curves \A and \B, we consider the third rotation curve, which is based on the terminal rotation curve data from \citep{2011MNRAS.414.2446M} with the assumption of a higher value $\Theta_{\circ}=239\kms$.

\subsection{The influence of the solar Galactocentric distance $R\sun$}
In our calculations we adopted the frequently used value of the solar Galactocentric distance $R\sun=8.0\,\mathrm{kpc}$, the same as in \citet{2005ARep...49..435A}, where a review of other estimations can be found. In modeling microlensing in the Galaxy, other values are sometimes preferred.
In the Galactic model presented by the EROS collaboration \citep{1999A&A...351...87E}, the value recommended by the IAU $R\sun=8.5\,\mathrm{kpc}$ is used. Because the integrand in the formula defining the optical depth \eqref{eq:depth} is small near the Sun's position, the choice of $R\sun$ cannot alter the results significantly, however. When $R\sun$ is increased from $R\sun=8.0\,\mathrm{kpc}$ to  $R\sun=8.5\,\mathrm{kpc}$ in the integral \eqref{eq:depth}, there is a corresponding change in the optical depth of $\Delta\tau=+0.1\times 10^{-6}$ at $b=1^{\circ}$ and only $\Delta\tau=-0.04\times 10^{-6}$ at $b=5^{\circ}$. One can also find a lower value for $R\sun$ of $7.62\,\mathrm{kpc}$ \citep{2005ApJ...628..246E}. Replacing $R\sun=8.0\,\mathrm{kpc}$ with  $R\sun=7.62\,\mathrm{kpc}$ in the integral \eqref{eq:depth}  gives rise to a change in the optical depth of $\Delta\tau=-0.1\times 10^{-6}$ at $b=1^{\circ}$ and only $\Delta\tau=+0.02\times 10^{-6}$ at $b=5^{\circ}$. This shows that the optical depth is not sensitive to the changes in the estimate of $R\sun$ within the limits most frequently met in the literature.

\subsection{The influence of the bulge  structure }

The disk model predicts a substitute surface density $\sigma(R)$ in the disk plane that accounts for the rotation curve. It can be regarded as the column mass density. If the true mass distribution is flattened then the description in terms
of the column mass density should be sufficient for a dynamical model, but an appropriate volume mass reconstruction is needed when studying the microlensing. In the previous section we assumed that the vertical profile of the total volume mass density $\rho(R,z)$ changed exponentially: $\rho(R,z)=\rho(R,0)\,e^{-\mid z\mid/h}=\sigma(R)\,e^{-\mid z\mid/h}/2h$.
In particular, the central bulge can be described in terms of its own column mass density, which contributes to the total mass density.
  But the observations of the bulge \citep[for example,][]{1997MNRAS.288..365B,2002MNRAS.330..591B} indicate that a spherically symmetric distribution of matter or even a three-axial ellipsoid would better approximate the central part of the Galaxy, and close to the center there can be some correction to the density profile we use. The stars residing in the central bulge contribute significantly to the optical depth, thus the particular geometry of mass distribution may have some effect on the accuracy of our result. Therefore,
 some estimation of the uncertainty is indispensable in our predictions concerning the optical depth.

 To this end, we replace the central part extending out to $1\kpc$ by a spherically symmetric bulge and check how this change in the mass distribution alters the optical depth. Since the dynamical mass inside $1\kpc$ is comparable in both models (the disk model gives $M_{1\kpc}=1.061\times10^{10}\msun$, whereas for the spherical bulge $M_{1\kpc}=1.137\times10^{10}\msun$), we do not expect large discrepancies in their predictions. A possible change by a factor of order of unity may result from a change in the iso-density surfaces and from a  small change
 in mass distribution outside $1\kpc$ compensating for any excess from the rotation that including a spherically symmetric part may have caused in the  gravitational potential.

\begin{figure}
\centering
\includegraphics[width=0.5\textwidth]{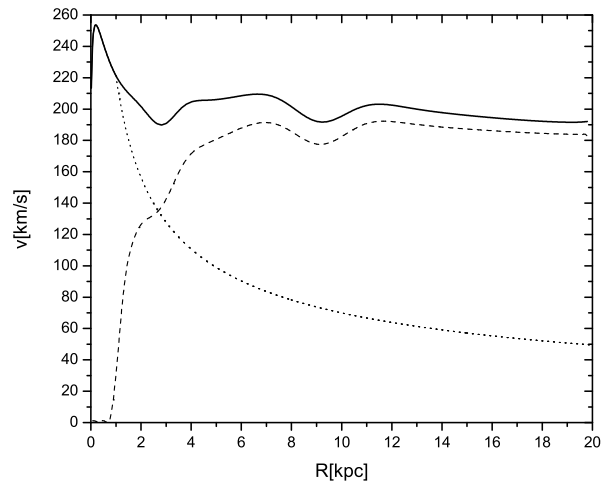}
\caption{\label{fig:milkibulge} The decomposition of the measured rotation curve into a bulge
and a disk contribution.}
\end{figure}

\begin{figure}
\centering
\includegraphics[angle=-90,width=0.5\textwidth]{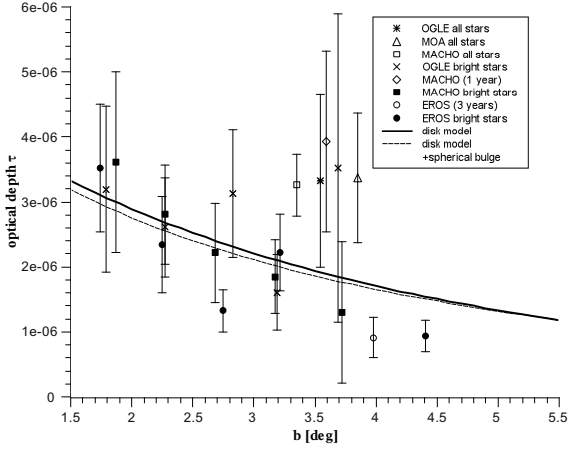}
\caption{The optical depth in a model with a spherically symmetric bulge [\emph{thin dashed line}] and the disk model prediction [\emph{solid line}]. \label{bulge_correction}}
\end{figure}

 With the bulge included, the integral for the optical depth splits as follows
\begin{eqnarray*}
\frac{c^2\,\tau_h(b)}{4\pi GR\sun^2}=\int_{0}^{s_b}s(1-s)\rho_{disk}(s)\dd{s}+\int_{s_b}^{1}s(1-s)\rho_{bulge}(s)\dd{s}.
\end{eqnarray*}
The first term is the contribution from the thick disk,
and the
second term is the contribution from the central bulge, where $\rho_{bulge}$ and $\rho_{disk}$ are understood as functions of $R(s)$ and $z(s)$ along the lines of sight.
The integration limits are determined by the bulge size $R_b=(1-s_b)R\sun=1\kpc$.
The density $\rho_{bulge}$ at the spherical radius $r(s)=\sqrt{R^2+z^2}=R\sun\sqrt{(1-s)^2+s^2b^2}$ can be found from the mass function $M(r)$ of the bulge
$$\rho_{bulge}(s)=\frac{1}{4\pi r^2(s)}\frac{\dd{M}}{\dd{r}}\Big\arrowvert_{r=r(s)},$$
and this $M(r)$ in turn
 contributes to the rotation curve.
We assume that the bulge accounts for the measured rotation curve for radii $R<R_b$. In this case we have
$M(r)=r\,v^2(r)/G$, while for $R > R_b$ we assume the contribution from the bulge to be Keplerian:  $v_b(R) = v(R_b)\sqrt{R_b/R}$.
As previously, for small $b$'s the disk volume density is $\rho_{disk}=\sigma_1(R\sun(1-s))\,\exp{-s|b|R\sun/h}/2h$, where
$\sigma_1$ accounts for the remaining part of rotation $v_1^2(R) = v^2(R)-v_b^2(R)$ not explained by the bulge \citep[a similar decomposition of rotation was applied to another galaxy in][]{acta}.
     The decomposition of the rotation
curve is plotted in fig.\ref{fig:milkibulge}.
The required surface density
$\sigma_1(R)$ corresponding to $v_1(R)$ can be found by applying the iteration method described in \cite{2008ApJ...679..373J}.
The resulting optical depth is plotted in fig.\ref{bulge_correction}.
As expected, the correction from the presence of the bulge is so small (of the order of $\Delta\tau=0.2\times10^{-6}$) that it cannot change the main conclusion of the previous paragraph.

\subsection{Influence of the vertical structure}
\begin{figure}
\includegraphics[angle=-90,width=0.5\textwidth]{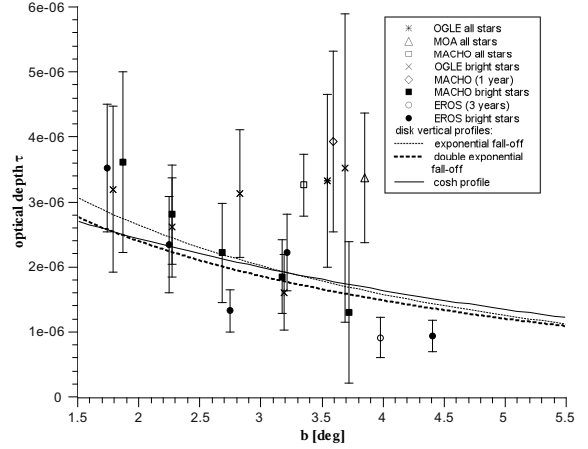}
\caption{The optical depth calculated after subtracting the gas contribution, with the inclusion of the spherical bulge, shown for all of the vertical density distributions considered in the text. The thin dotted line represents the reference model -- a disk with the exponential vertical profile with the scale height $h=325\,\mathrm{pc}$. The thick dotted line corresponds to the disk with a double exponential vertical profile, and the solid line represents the hyperbolic cosine vertical profile.\label{fig_vertical}  }
\end{figure}

The disk model predicts only the column mass density $\sigma(R)$, and its reprojection to a volume mass density $\rho(R,z)$ depends on the additional assumption about the adopted  density fall-off in the vertical direction.
It is therefore indispensable to estimate the influence of this assumption on the predicted optical depth.
 As the starting point, we adopt a single exponential vertical fall-off $\rho(R,z)=\rho(R,0)\,\mathrm{e}^{-\mid z\mid/h}$ with the scale height parameter $h=325\,\mathrm{pc}$. The dependence on the scale height parameter has been already checked and presented in figure \ref{tau_h}. It shows that when the effective change of the scale height is not large, the results cannot be changed significantly, especially for the higher latitudes. In this section, we compare our result with the ones for two other vertical distributions of matter. First, we consider a double exponential vertical profile:
$$\rho_{disk}(R,z)=\rho(R,0)\left((1-\beta)\,\mathrm{e}^{-\mid z\mid/h_1}+\beta\,\mathrm{e}^{-\mid z\mid/h_2}\right).$$
The parameters $h_1=320\,\mathrm{pc}$ for the thinner disk, $h_2=643\,\mathrm{pc}$ for the thicker disk, and the fraction $\beta=0.216$ are taken from \cite{1997ApJ...482..913G}. Likewise, we also consider the \emph{hyperbolic cosine} profile used in \cite{2004IAUS..220..201V}:
$$\rho_{disk}(R,z)=\rho(R,0)\,\frac{1}{\mathrm{cosh}(z/h)}. $$
As we show in section \ref{sec:mtol}, this profile approximates the observations better in the solar neighborhood and improves the mass-to-light ratio. The comparison of the model  curves $\tau(b)$ for the disk vertical density distribution profiles mentioned above is presented in figure \ref{fig_vertical}. The calculations are made after subtracting the gas contribution and by considering the spherical bulge component.
As we expected, the choice of the particular vertical density fall-off could only slightly change the results. Replacement of a single exponential profile by a double exponential profile reduces the optical depth approximately by $\Delta\tau=0.2\times10^{-6}$.

Finally, we discuss a more complicated situation.
The case of a three-axial ellipsoid is beyond the scope of the simple axi-symmetric dynamical model of the Galaxy, nevertheless, a correction to the optical depth connected with the presence of the bar should be within an order of magnitude as in the case of the spherically symmetric bulge considered earlier. We have seen above that our result is very stable against small changes in the mass distribution. Similarly, considering a bar-like structure would only introduce higher multipoles to the mass distribution that should not affect qualitatively our main conclusion. In the next section we study this issue in a more detail.

\subsection{Model uncertainties and the bar issue}\label{section_bar}

It is plausible that very particular assumptions about the dynamical model of the Galaxy could affect the resulting optical depth.
If the model is based on the rotation curve, in preparation of which the axial symmetry is assumed, it is natural to assume axisymmetry for the distribution of the sources of light.
This symmetry is consistent with the disk symmetry and with the spherical symmetry, whereas it is not consistent with the three-axial symmetry of a bar-like structure. Because of this, an axisymmetric model is not adequate for estimating the optical depth in a single, particular direction. That is why we are using a $\tau(b)$ observable, which has the meaning of the optical depth over slices of fixed longitude with latitudes around a particular $b$. Taking the statistical character of $\tau(b)$ into account, we performed a Monte Carlo simulation that mimics the observational situation of detecting microlensing events from many sources with various longitudes and various distances to the observer. Although the longitude of each field is well determined, the distance to the particular source of light (described by the $\chi$ parameter) is practically unknown. We have chosen a sample of pairs $(l,\chi)$ for which we used a probability distribution weighted by the surface mass density $\sigma$. The result for the $\tau(b)$ calculated for this sample is depicted in figure 6. It shows that on average the approximation for which the sources are located on the symmetry axis works quite well.

One of the advantages of the Monte Carlo approach is that it enables one to estimate a possible change or uncertainty in the determination of $\tau$ that could be introduced by deformations of the central part, in particular those due to the presence of the bar. As previously, we do not expect changes in the main conclusion. This remark is suggested by the fact that the dependence of the optical depth  on the longitude $\tau(l)$ (within latitudinal slices of longitudes from around a particular $l$) is much weaker than the dependence on the latitude $\tau(b)$. According to \cite{2006A&A...454..185H}, the optical depth changes by about $\Delta\tau=1.3\times10^{-6}$ in the range $l\in(-5^\circ,5^\circ)$ while the optical depth change in the considered range of latitudes is about $\Delta\tau=2.7\times10^{-6}$. Our disk model predicts the change in the optical depth in the range $l\in(-5^\circ,5^\circ)$ of about $\Delta\tau=1.8\times10^{-6}$. Then, the correction connected with the inclusion of the bar should be less than $\Delta\tau=0.5\times10^{-6}$. To estimate the influence of the bar on our results, we employ the three-axial G2 model developed in \cite{1995ApJ...445..716D}. It gives a volume emissivity of the sources of light in the Galactic center:
\begin{equation}\label{emissivity}
\eta(x',y',z')=\eta_0\,\mathrm{exp}\left[-\frac{1}{2}\sqrt{\left[\left(\frac{x'}{x_{0}}\right)^2+\left(\frac{y'}{y_{0}}\right)^2 \right]^2+\frac{z'^4}{z_{0}{}^4}}\right].
\end{equation}
In section \ref{sec:appl} we used the following notation for the cartesian coordinates of the source $X_{\otimes}=[x,y,z]=R\sun\,[1-(1+\chi)\cos b\cos l\,,\,-(1+\chi)\cos b\sin l\,,\,(1+\chi)\sin b]$, which differs from the naming convention used in \cite{1995ApJ...445..716D}. If we denote by
$x'$, $y'$, $z'$ as the cartesian coordinates of their coordinate system, then
the  appropriate transformation reads as
 $x'\to y$, $y'\to-x$, and $z'\to z$.

 We applied the formula (\ref{emissivity}) to perform our Monte Carlo simulation in which the sources of light are chosen randomly with the probability distribution proportional to the volume emissivity. This imitates the situation in which the sources are distributed within the three-axial bar. We take the parameters $x_{0}=2.01\,\mathrm{kpc}$, $y_{0}=0.62\,\mathrm{kpc}$, and $z_{0}=0.44\,\mathrm{kpc}$ fitted in the $3,5\,\mu\mathrm{m}$ band, although the values obtained in other bands give roughly the same result. In figure \ref{fig10} below, we present the comparison of the Monte Carlo simulations assuming the axial symmetry and one that considers the G2 three-axial model. For this three-axial model, an extended range for longitudes is assumed $\mid l\mid<0.085$ (that is, $\pm5^\circ$), which covers all the fields used in the preparation of data  \citep[see figure 14 in ][]{2006A&A...454..185H}. In the central part of the bulge (say, $\pm1^\circ$ in the longitude), there is practically no difference between the predictions of the three-axial model and the spherically symmetric bulge models.
\begin{figure}
\includegraphics[angle=-90,width=0.5\textwidth]{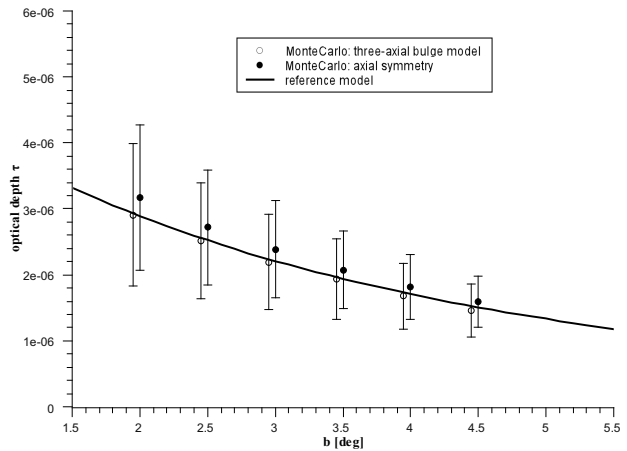}
\caption{\label{fig10} A Monte Carlo simulation of the optical depth (and the standard deviation) in the axisymmetric case [\emph{solid circles and bars}] and in the presence of a three-axial bar [\emph{open circles and bars}]. The solid line represents the model curve based on the integral \eqref{eq:depth_full}, in which case the sources of light lie on the symmetry axis.}
\end{figure}
One can see that the results for the sources distributed within the three-axial bulge are also consistent with the main disk model prediction. The correction due to the bar, combined with the dependence on the choice of the field range, gives no more than $\Delta\tau=0.4\times10^{-6}$. We neglect that the non-axisymmetric component affects the dynamics of the inner part of the Galaxy a bit. However, this correction also cannot be large, because the photometrically determined bulge mass in the G2 model $M_{G2}=1.3\times 10^{10}\,M_{\odot}$ \cite{1995ApJ...445..716D} is comparable to our estimation $M_{1kpc}=1.137\times 10^{10}\,M_{\odot}$.

\subsection{The hidden mass estimation}
\begin{figure}
\includegraphics[angle=-90,scale=0.5]{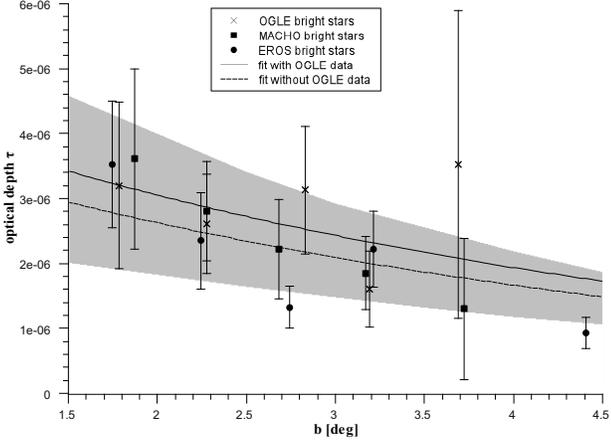}
\caption{\label{fig_fitting} The $(1-f)\,c_1\mathrm{exp}(-c_2\mid b\mid)\times10^{-6}$ fit to the OGLE, MACHO, and EROS bright stars optical depth data [\emph{solid line}] and a similar fit to the narrowed data of MACHO and EROS bright stars data [\emph{dashed line}]. The filled area corresponds to the standard deviation range of the Monte Carlo simulation (including bar) described in the previous section. The $c_1\mathrm{exp}(-c_2\mid b\mid)\times10^{-6}$ curve is not plotted for clarity. }
\end{figure}

To estimate the amount of the hidden mass we must take the $\tau(b)$ prediction in the thin disk model framework into account after subtraction of the gas contribution, using  rotation curve \B  (which gives a more massive disk than rotation curve \A), considering the spherical bulge component and the disk double--exponential vertical fall-off. The result can be parameterized as follows: $\tau(b)=c_1\mathrm{exp}(-c_2\mid b\mid)\times10^{-6}$ with coefficients $c_1=4.23\pm0.65$ and $c_2=0.228\pm0.056$. Not only the neutral and molecular hydrogen and helium, but also other components of the total dynamical mass inside the solar orbit of $M_{tot}=6.17\times 10^{10}M_{\odot}$ contribute to this result. A fraction of $M_{tot}$ could consist of matter that is neither in the form of compact objects nor in the form of hydrogen and helium clouds, which have  already been subtracted. Because it would not contribute to the measured optical depth, we call it the hidden mass. It overestimates the model optical depth $\tau(b)$ by a factor $\tau_{hidden}(b)$, which can be obtained by inserting the volume mass density of the hidden mass $\rho_{hidden}(R,z)$ into the integral \eqref{eq:depth_full}. Unfortunately, the $\rho_{hidden}(R,z)$ profile is not known. Since this profile appears in the integrand of the relevant formula, there are various possible profiles of $\rho_{hidden}(R,z)$ that could lead to the optical depth correction $\tau_{hidden}(b)$ giving $\tau(b)-\tau_{hidden}(b)$ as the best fit to the observational optical-depth data points. For the simplicity of the estimation, we may assume that the hidden mass density is a constant fraction of the dynamical mass density (without gas) $\rho_{hidden}(R,z)=f\,\rho(R,z)$. Then, the correction factor is $\tau_{hidden}(b)=f\,\tau(b)$, and the observed optical depth is expected to be $(1-f)\tau(b)$. We find the fraction $f$ by fitting the formula $(1-f)\,c_1\mathrm{exp}(-c_2\mid b\mid)\times10^{-6}$ to the bright stars $\tau$ observations, by means of the least squares method weighted by the observational uncertainties. The results of the fitting procedure are depicted in figure \ref{fig_fitting}.
Taking the OGLE, MACHO, and EROS data into account, we obtain $f=-0.139\pm0.059$.  Because this fraction is negative, that would mean that an additional dynamical mass of about $\mid f\,M_{tot}\mid=(0.86\pm0.36)\times 10^{10}M_{\odot}$ is needed. The data points collected by the OGLE collaboration lie above the points collected by other groups and have a bit larger uncertainties. If we restricted the analysis to the MACHO and EROS data, we would get $f=0.018\pm0.067$, which translates to $\mid f\,M_{tot}\mid=(0.11\pm0.41)\times 10^{10}M_{\odot}$ of the hidden mass. One should regard all this as rough estimates. The two fits that we present are only used to find the expectation value of the optical depth $E[\tau(b)]$ by means of the least squares method. On the other hand, we know what the expected variance of the optical depth predicted by the thin disk model is, because we have performed the Monte Carlo simulation, which gives us the uncertainties $\Delta\tau_{MonteCarlo}(b)$. They could differ from the fit residuals. The variance assigned to our model should be $\mathrm{var}[\tau(b)]=(\Delta\tau_{MonteCarlo}(b))^2$. Instead of checking the goodness of the fit, one can compare the model distribution $\{E[\tau(b)],\mathrm{var}[\tau(b)]\}$ with the observational one $\{\bar{\tau}(b),\sigma^2(b)\}$. The average $\bar{\tau}(b)$ can be taken within some narrow band in the latitude $b\pm0.2\mathrm{deg}$, and the $\sigma(b)$ is the observational uncertainty within the band. To check the statistical hypothesis that these two distributions have the same expectation values $E[\tau(b)]=\bar{\tau}(b)$, one should compute the quantity:
$$z_b=\frac{E[\tau(b)]-\bar{\tau}(b)}{\sqrt{\frac{\mathrm{var}[\tau(b)]}{n}+
\frac{\sigma^2(b)}{n_b}}}.$$
Here $n=10000$ denotes the number of points generated in the Monte Carlo simulation for the fixed latitude $b$, and $n_b$ is the number of the observational points within the band $b\pm0.2\mathrm{deg}$. The variable $z_b$ has the normal distribution $N(0,1)$. Below, in Tables \ref{tab1} and \ref{tab2}, the appropriate values are listed in the case without the OGLE data and in the case in which all the data are considered.

\begin{table}
\begin{tabular}
{|c|c|c|c|c|c|}\hline
b  & $E[\tau(b)]$ & $\mathrm{var}[\tau(b)]$ & $\bar{\tau}(b)$ & $\sigma^2(b)$ & $z_b$\\
$[deg]$ & $[10^{-6}]$ & $[10^{-12}]$ & $[10^{-6}]$ & $[10^{-12}]$ & $ $\\
\hline
$1.8$   & $2.75$                    & $2.17$                                & $3.56$                       & $1.40$                      & $-0.97$ \\
$2.2$   & $2.49$                    & $1.44$                                & $2.57$                       & $0.56$                      & $-0.15$ \\
$2.7$   & $2.24$                    & $1.44$                                & $1.77$                       & $0.29$                      & $1.21$ \\
$3.2$   & $2.00$                    & $0.99$                                & $2.03$                       & $0.33$                      & $-0.07$ \\
$3.7$   & $1.78$                    & $0.66$                                & $1.30$                       & $1.18$                      & $0.44$ \\
\hline
\end{tabular}
\caption{\label{tab1}The expectation value, the variance, and the test statistics $z_b$ in the case without OGLE data points.   }
\end{table}
\begin{table}
\begin{tabular}
{|c|c|c|c|c|c|}\hline
b  & $E[\tau(b)]$ & $\mathrm{var}[\tau(b)]$ & $\bar{\tau}(b)$ & $\sigma^2(b)$ & $z_b$\\
$[deg]$ & $[10^{-6}]$ & $[10^{-12}]$ & $[10^{-6}]$ & $[10^{-12}]$ & $ $\\
\hline
$1.8$   & $3.20$                    & $2.17$                                & $3.44$                       & $1.48$                      & $-0.35$ \\
$2.2$   & $2.92$                    & $1.44$                                & $2.59$                       & $0.57$                      & $0.76$ \\
$2.7$   & $2.60$                    & $1.44$                                & $2.22$                       & $0.47$                      & $0.94$ \\
$3.2$   & $2.33$                    & $0.99$                                & $1.89$                       & $0.33$                      & $1.30$ \\
$3.7$   & $2.07$                    & $0.66$                                & $2.41$                       & $2.98$                      & $-0.28$ \\
\hline
\end{tabular}
\caption{\label{tab2}The expectation value, the variance, and the test statistics $z_b$ in the case where all data are considered   }
\end{table}

In both cases the test statistics $z_b$ have values relatively close to zero. On a reasonable confidence level of $p=0.01$ one cannot reject the null hypothesis that the model distribution $\{E[\tau(b)],\mathrm{var}[\tau(b)]\}$ and the observational distribution $\{\bar{\tau}(b),\sigma^2(b)\}$ have the same expectation values. By taking the highest value $z_b=1.3$, one could reject this hypothesis on the confidence level $p=0.19$; however, this would lead to a very high probability $p=0.19$ of the statistical error of the first kind (of the rejection of the true null hypothesis). With a smaller number $n$, the agreement is even better (because for $n=10000$ the term $\mathrm{var}[\tau(b)]/n$ can be neglected).

In other words, almost all observational data lie close to the expectation value $E[\tau(b)]$, within the $\pm\Delta_{MonteCarlo}(b)$ range, which is depicted in figure \ref{fig_fitting}. Since the fits we have obtained are close to the value $f=0$, and the observational data are consistent with the Monte Carlo simulation, we are led to the conclusion that all of the dynamical mass corresponding to the rotation curve in the disk model (and after subtraction of gas) is consistent with the optical depth. The amount of the hidden mass or an additional dynamical mass that could be needed to account for the optical depth observation, is negligible in the $0<R<R\sun$ region.

\subsection{\label{sec:mtol}The mass-to-light ratio in the solar neighborhood in the disk model}
As a cross-check for the dynamical model of the Galaxy one can ask about its prediction of the light-to-mass ratio in the solar neighborhood.
To determine the profile of the local mass-to-light ratio one needs a brightness profile. It is
difficult to determine the brightness profile for Galaxy in the form of a single curve. There were only attempts to determine the overall brightness in the solar vicinity by direct counting of nearby stars. Such brightness estimate in our neighborhood is quite accurate (but questionable as globally representative for the Galaxy at the same radial distance).
In \cite{2006MNRAS.372.1149F} the local I-band brightness is $29.54 \lsun/\pc^2$. It is determined based on star counts out to $200\pc$ above the Galactic midplane. In the same volume there is $13.2 \msun/\pc^2$ of hydrogen $(HI+HII+H_2)$.
Given these data, we obtain $110\msun/\pc^2$  or $140 \msun/\pc^2$ in the disk model for the column mass density corresponding to rotation curves \A or \B, respectively. Subtracting four thirds of hydrogen (thus including $He$ from the bariogenesis), because the gas is nonluminous in this band, we obtain $M/L_I=3.13$ or $M/L_I=4.14$, respectively, in the solar vicinity. Although overestimated, these values are low (and could be still reduced by including more distant objects, above $z=200\pc$, which have not been  included in the counting).

For a better estimate of the mass-to-light ratio in the solar neighborhood we can test other models of distribution of stars above the midplane.
In \cite{2004IAUS..220..201V}, the profile ansatz $1/\cosh(z/h)$ was assumed as approximating the observations better than the usual exponential falloff.
Assuming our total (column) surface density of $110\msun/\pc^2$ and $h=330\pc$  we obtain $\rho=0.11\msun/\pc^3$ in the solar vicinity, so that the integrated density out to  $h=200\pc$ gives $42\msun/\pc^2$ and the corresponding $M/L_I=0.83$ or  $M/L_I=1.1$ for rotation curves A or B, respectively.
With exponential vertical falloff $\exp{-|z|/h}$  (with $h=350\pc$), we would have $M/L_I=0.95$ or  $M/L_I=1.47$ for rotation curves \A or \B, respectively.
 These are quite reasonable and low local mass-to-light ratios, which seem to be compatible with the recent findings from the star kinematics in the solar neighborhood \citep{Bidin}.

\section{Summary and conclusions}
The microlensing method determines the optical depth observable. The observable can be modeled provided the volume mass density of compact objects is known.  We assumed that the total volume mass density has an exponential vertical falloff and that the associated column mass density accounts for the tangential component of rotation  in the thin disk model approximation. The structure of the central bulge and the presence of the bulge do not alter the results significantly, so our results are also insensitive to slight changes in the rotation curves such as the choice of the parameter $\Theta$.) Having subtracted the gas contribution from the volume mass density obtained in this way, we modeled the optical depth and then compared the model with the optical depth measurements. It follows from the analysis that the amount of mass seen through gravitational microlensing in the region $0<R<R\sun$ is consistent with our mass model. So that our model of the optical depth conform more to realistic measurements, we carried out a Monte Carlo simulation. It proved to mimic the measurements quite satisfactorily, since on the reasonable confidence level one could not reject the hypothesis that its expectation value equaled the expectation value of the observational data. We estimated the limits for the amount of nonbaryonic dark matter in our model inside radius $R\sun$ to be  $(0.11\pm0.4)\times10^{10}\msun$.
Since microlensing only detects compact objects, the result suggests that the nonbaryonic mass component may be negligible in this region because the mass of gas and that of compact objects seen through microlensing suffice to account for Galaxy rotation in this region.

\bibliography{lensing}

\begin{thebibliography}{}

\bibitem[\protect\citeauthoryear{{Alard}}{{Alard}}{1997}]{1997A&A...321..424A}
{Alard} C.,  1997, \aap, 321, 424

\bibitem[\protect\citeauthoryear{{Avedisova}}{{Avedisova}}{2005}]{2005ARep...49..435A}
{Avedisova} V.~S.,  2005, Astronomy Reports, 49, 435

\bibitem[\protect\citeauthoryear{{Binney}, {Gerhard} \& {Spergel}}{{Binney}
  et~al.}{1997}]{1997MNRAS.288..365B}
{Binney} J.,  {Gerhard} O.,    {Spergel} D.,  1997, \mnras, 288, 365

\bibitem[\protect\citeauthoryear{{Bissantz} \& {Gerhard}}{{Bissantz} \&
  {Gerhard}}{2002}]{2002MNRAS.330..591B}
{Bissantz} N.,  {Gerhard} O.,  2002, \mnras, 330, 591

\bibitem[\protect\citeauthoryear{{Blitz}, {Fich} \& {Stark}}{{Blitz}
  et~al.}{1982}]{1982ApJS...49..183B}
{Blitz} L.,  {Fich} M.,    {Stark} A.~A.,  1982, \apjs, 49, 183

\bibitem[\protect\citeauthoryear{{Bovy} \& {Tremaine}}{{Bovy} \&
  {Tremaine}}{2012}]{2012arXiv1205.4033B}
{Bovy} J.,  {Tremaine} S.,  2012, \apj, 756, 89

\bibitem[\protect\citeauthoryear{{Bratek}, {Ja{\l}ocha} \&
  {Kutschera}}{{Bratek} et~al.}{2008}]{2008MNRAS.391.1373B}
{Bratek} {\L}.,  {Ja{\l}ocha} J.,    {Kutschera} M.,  2008, \mnras, 391, 1373

\bibitem[\protect\citeauthoryear{{Burton} \& {Gordon}}{{Burton} \&
  {Gordon}}{1978}]{1978A&A....63....7B}
{Burton} W.~B.,  {Gordon} M.~A.,  1978, \aap, 63, 7

\bibitem[\protect\citeauthoryear{{Clemens}}{{Clemens}}{1985}]{1985ApJ...295..422C}
{Clemens} D.~P.,  1985, \apj, 295, 422

\bibitem[\protect\citeauthoryear{{Demers} \& {Battinelli}}{{Demers} \&
  {Battinelli}}{2007}]{2007A&A...473..143D}
{Demers} S.,  {Battinelli} P.,  2007, \aap, 473, 143

\bibitem[\protect\citeauthoryear{{Derue} \& at al}{{Derue} \&
  at~al}{1999}]{1999A&A...351...87E}
{Derue} F.,  at al 1999, \aap, 351, 87

\bibitem[\protect\citeauthoryear{{Dwek}, {Arendt}, {Hauser}, {Kelsall},
  {Lisse}, {Moseley}, {Silverberg}, {Sodroski} \& {Weiland}}{{Dwek}
  et~al.}{1995}]{1995ApJ...445..716D}
{Dwek} E.,  {Arendt} R.~G.,  {Hauser} M.~G.,  {Kelsall} T.,  {Lisse} C.~M.,
  {Moseley} S.~H.,  {Silverberg} R.~F.,  {Sodroski} T.~J.,    {Weiland} J.~L.,
  1995, \apj, 445, 716

\bibitem[\protect\citeauthoryear{{Eisenhauer}, {Genzel} \& at al}{{Eisenhauer}
  et~al.}{2005}]{2005ApJ...628..246E}
{Eisenhauer} F.,  {Genzel}   at al 2005, \apj, 628, 246

\bibitem[\protect\citeauthoryear{{Fich}, {Blitz} \& {Stark}}{{Fich}
  et~al.}{1989}]{1989ApJ...342..272F}
{Fich} M.,  {Blitz} L.,    {Stark} A.~A.,  1989, \apj, 342, 272

\bibitem[\protect\citeauthoryear{{Flynn}, {Holmberg}, {Portinari}, {Fuchs} \&
  {Jahrei{\ss}}}{{Flynn} et~al.}{2006}]{2006MNRAS.372.1149F}
{Flynn} C.,  {Holmberg} J.,  {Portinari} L.,  {Fuchs} B.,    {Jahrei{\ss}} H.,
  2006, \mnras, 372, 1149

\bibitem[\protect\citeauthoryear{{Gould}, {Bahcall} \& {Flynn}}{{Gould}
  et~al.}{1997}]{1997ApJ...482..913G}
{Gould} A.,  {Bahcall} J.~N.,    {Flynn} C.,  1997, \apj, 482, 913

\bibitem[\protect\citeauthoryear{{Gradshtein}, {Ryzhik}, {Jeffrey} \&
  {Zwillinger}}{{Gradshtein} et~al.}{2007}]{Ryzhik}
{Gradshtein} I.,  {Ryzhik} I.,  {Jeffrey} A.,    {Zwillinger} D.,  2007, Table
  of integrals, series and products.
Academic Press

\bibitem[\protect\citeauthoryear{{Hamadache}, {} \& {et al.}}{{Hamadache}
  et~al.}{2006}]{2006A&A...454..185H}
{Hamadache} C.,  {}   {et al.} 2006, \aap, 454, 185

\bibitem[\protect\citeauthoryear{{Honma} \& {Sofue}}{{Honma} \&
  {Sofue}}{1997}]{1997PASJ...49..453H}
{Honma} M.,  {Sofue} Y.,  1997, \pasj, 49, 453

\bibitem[\protect\citeauthoryear{{Ja{\l}ocha}, {Bratek} \&
  {Kutschera}}{{Ja{\l}ocha} et~al.}{2008}]{2008ApJ...679..373J}
{Ja{\l}ocha} J.,  {Bratek} {\L}.,    {Kutschera} M.,  2008, \apj, 679, 373

\bibitem[\protect\citeauthoryear{{Ja{\l}ocha}, {Bratek} \&
  {Kutschera}}{{Ja{\l}ocha} et~al.}{2010}]{acta}
{Ja{\l}ocha} J.,  {Bratek} {\L}.,    {Kutschera} M.,  2010, \appb, 41, 1383

\bibitem[\protect\citeauthoryear{{Ja{\l}ocha}, {Bratek}, {Kutschera} \&
  {Skindzier}}{{Ja{\l}ocha} et~al.}{2010a}]{2010MNRAS.406.2805J}
{Ja{\l}ocha} J.,  {Bratek} {\L}.,  {Kutschera} M.,    {Skindzier} P.,  2010a,
  \mnras, 406, 2805

\bibitem[\protect\citeauthoryear{{Ja{\l}ocha}, {Bratek}, {Kutschera} \&
  {Skindzier}}{{Ja{\l}ocha} et~al.}{2010b}]{2010MNRAS.407.1689J}
{Ja{\l}ocha} J.,  {Bratek} {\L}.,  {Kutschera} M.,    {Skindzier} P.,  2010b,
  \mnras, 407, 1689

\bibitem[\protect\citeauthoryear{{McMillan}}{{McMillan}}{2011}]{2011MNRAS.414.2446M}
{McMillan} P.~J.,  2011, \mnras, 414, 2446

\bibitem[\protect\citeauthoryear{{Misiriotis}, {Xilouris}, {Papamastorakis},
  {Boumis} \& {Goudis}}{{Misiriotis} et~al.}{2006}]{2006A&A...459..113M}
{Misiriotis} A.,  {Xilouris} E.~M.,  {Papamastorakis} J.,  {Boumis} P.,
  {Goudis} C.~D.,  2006, \aap, 459, 113

\bibitem[\protect\citeauthoryear{{Moni Bidin}, {Carraro}, {M{\'e}ndez} \&
  {Smith}}{{Moni Bidin} et~al.}{2012}]{Bidin}
{Moni Bidin} C.,  {Carraro} G.,  {M{\'e}ndez} R.~A.,    {Smith} R.,  2012,
  \apj, 751, 30

\bibitem[\protect\citeauthoryear{{Moniez}}{{Moniez}}{2010}]{2010GReGr..42.2047M}
{Moniez} M.,  2010, General Relativity and Gravitation, 42, 2047

\bibitem[\protect\citeauthoryear{{Paczynski}}{{Paczynski}}{1986}]{1986ApJ...304....1P}
{Paczynski} B.,  1986, \apj, 304, 1

\bibitem[\protect\citeauthoryear{{Paczynski}}{{Paczynski}}{1996}]{1996ARA&A..34..419P}
{Paczynski} B.,  1996, \araa, 34, 419

\bibitem[\protect\citeauthoryear{{Popowski}, {at al} \& {MACHO
  Collaboration}}{{Popowski} et~al.}{2005}]{2005ApJ...631..879P}
{Popowski} P.,  {at al}   {MACHO Collaboration} 2005, \apj, 631, 879

\bibitem[\protect\citeauthoryear{{Schneider}, {Kochanek} \&
  {Wambsganss}}{{Schneider} et~al.}{2006}]{schneider2006gravitational}
{Schneider} P.,  {Kochanek} C.,    {Wambsganss} J.,  2006, Gravitational
  lensing: strong, weak and micro.
Saas-Fee Advanced Course: Swiss Society for Astrophysics and Astronomy,
  Springer

\bibitem[\protect\citeauthoryear{{Smith}, {Wo{\'z}niak}, {Mao} \&
  {Sumi}}{{Smith} et~al.}{2007}]{2007MNRAS.380..805S}
{Smith} M.~C.,  {Wo{\'z}niak} P.,  {Mao} S.,    {Sumi} T.,  2007, \mnras, 380,
  805

\bibitem[\protect\citeauthoryear{{Sofue}, {Honma} \& {Omodaka}}{{Sofue}
  et~al.}{2009}]{2009PASJ...61..227S}
{Sofue} Y.,  {Honma} M.,    {Omodaka} T.,  2009, \pasj, 61, 227

\bibitem[\protect\citeauthoryear{{Sofue}, {Tutui}, {Honma}, {Tomita},
  {Takamiya}, {Koda} \& {Takeda}}{{Sofue} et~al.}{1999}]{1999ApJ...523..136S}
{Sofue} Y.,  {Tutui} Y.,  {Honma} M.,  {Tomita} A.,  {Takamiya} T.,  {Koda} J.,
     {Takeda} Y.,  1999, \apj, 523, 136

\bibitem[\protect\citeauthoryear{{Sumi} \& {at al}}{{Sumi} \& {at
  al}}{2003}]{2003ApJ...591..204S}
{Sumi} T.,  {at al} 2003, \apj, 591, 204

\bibitem[\protect\citeauthoryear{{Sumi}, {Wo{\'z}niak}, {Udalski},
  {Szyma{\'n}ski}, {Kubiak}, {Pietrzy{\'n}ski}, {Soszy{\'n}ski},
  {{\.Z}ebru{\'n}}, {Szewczyk}, {Wyrzykowski} \& {Paczy{\'n}ski}}{{Sumi}
  et~al.}{2006}]{2006ApJ...636..240S}
{Sumi} T.,  {Wo{\'z}niak} P.~R.,  {Udalski} A.,  {Szyma{\'n}ski} M.,  {Kubiak}
  M.,  {Pietrzy{\'n}ski} G.,  {Soszy{\'n}ski} I.,  {{\.Z}ebru{\'n}} K.,
  {Szewczyk} O.,  {Wyrzykowski} {\L}.,    {Paczy{\'n}ski} B.,  2006, \apj, 636,
  240

\bibitem[\protect\citeauthoryear{{van Altena}, {Korchagin}, {Girard}, {Dinescu}
  \& {Borkova}}{{van Altena} et~al.}{2004}]{2004IAUS..220..201V}
{van Altena} W.~F.,  {Korchagin} V.~I.,  {Girard} T.~M.,  {Dinescu} D.~I.,
  {Borkova} T.~V.,  2004, in {S.~Ryder, D.~Pisano, M.~Walker, \& K.~Freeman}
  ed., Dark Matter in Galaxies Vol.~220 of IAU Symposium, {Galactic Disk
  Surface Density in the Solar Neighbourhood}.
p.~201

\end{thebibliography}
\bibliographystyle{mn2e}

\appendix
\section{The integral transforms appearing in the disk model}\label{appendix_A}
The relation (\ref{eq:sigmamoja}) used in the text can be obtained from an equivalent integral expression derived in \citep{2008MNRAS.391.1373B} \begin{eqnarray*}
\sigma(R)=
\frac{1}{\pi^2G}  \left[\int\limits_0^R
v^2(\chi)\biggl(\frac{K\br{{\chi}/{R}}}{ R\
\chi}-\frac{R}{\chi}
\frac{E\br{{\chi}/{R}}}{R^2-\chi^2}\biggr)\ud{\chi}
\cdots\right. \\ \left. \cdots+\int\limits_R^{\infty}v^2(\chi)
\frac{E\br{{R}/{\chi}}
}{\chi^2-R^2}\,\ud{\chi}\right].
\end{eqnarray*}  To this end one can use the identities $2E\br{k}=\br{1+k}E\br{\mu_k}+\br{1-k}K\br{\mu_k}$ and
$\br{1+k}K\br{k}=K\br{\mu_k}$ with $\mu_k=\frac{2\sqrt{k}}{1+k}$, applied separately for  $\chi<R$ (with $k=\chi/R$) and for $R<\chi$ (with $k=R/\chi$).
 The inverse relation mapping the surface density to the rotation law can be found directly from the definition of the gravitational potential, without the usual intermediate step through Henkel transforms\footnote{The gravitational potential due to axisymmetric thin disk with surface mass density $\sigma$ in cylindrical coordinates is $$\Phi\br{R,z}=-4G\int\limits_{0}^{\infty}
\frac{\chi \sigma(\chi)\ud{\chi}}{\sqrt{(R+\chi)^2+z^2}}\int_{0}^{\pi/2}\frac{\ud{\alpha}}{\sqrt{1-\frac{4R\chi}{
(r+\chi^2)+z^2}\sin^2\br{\alpha}}
}$$ and the definition of elliptic function $K$ can be used. Differentiation with respect to $R$ and taking the limit $z\to0$ (the velocity on circular orbits in the disk plane is $v^2(R)=R\,\partial_R\Phi\br{R,0}$) gives the desired result.} (note the sign difference in \eqref{eq:sigmamoja})
\begin{equation}\label{eq:inv}\frac{v^2\br{R}}{R}=2G\int\limits_{0}^{\infty}
\sq{
\frac{K\!\br{\mu_x}}{1+x}+\frac{E\!\br{\mu_x}}{1-x}}\sigma\!\br{R\,x}x\,\ud{x}.
\end{equation} The two integrals \eqref{eq:sigmamoja} and \eqref{eq:inv} are quite symmetric, the inversion transform $x\to1/x$ leaving $\mu_x$ invariant links the integral kernels of the two integrals. We see that two physical quantities, the disk surface density $\sigma(R)$ and the centrifugal acceleration on a circular orbit $\frac{v^2(R)}{R}$ in the disk plane, are integral transforms of each other.  It would be interesting to investigate this striking property of the disk model in more detail in the context of the theory of integral transforms.

\end{document}